\documentclass[twocolumn]{aastex62}
\usepackage{amsmath,amstext}
\usepackage{comment}
\usepackage{apjfonts} 
\usepackage{chngcntr}
\usepackage{float}
\usepackage{lineno}

\newcommand{\be}{\begin{eqnarray}}
\newcommand{\ee}{\end{eqnarray}}
\newcommand{\lp}{\left(}
\newcommand{\rp}{\right)}

\begin{document}

\normalsize
\interfootnotelinepenalty=10000


\title{\textbf{Estimating Ejecta Masses of Stripped Envelope Supernovae Using Late-Time Light Curves}}

\author{Annastasia Haynie}
\affiliation{Department of Physics and Astronomy, University of Southern California, Los Angeles, CA 90089, USA; ahaynie@usc.edu}
\affiliation{The Observatories of the Carnegie Institution for Science, 813 Santa Barbara St., Pasadena, CA 91101, USA}

\author{Anthony L. Piro}
\affiliation{The Observatories of the Carnegie Institution for Science, 813 Santa Barbara St., Pasadena, CA 91101, USA}

\begin{abstract}
    Stripped-envelope supernovae (SESNe) are a subclass of core-collapse supernovae that are deficient in hydrogen (SN~IIb, SN~Ib) and possibly helium (SN~Ic) in their spectra. Their progenitors are likely stripped of this material through a combination of stellar winds and interactions with a close binary companion, but the exact ejecta mass ranges covered by each subtype and how it relates to the zero-age main-sequence progenitor mass is still unclear. Using a combination of semi-analytic modeling and numerical simulations, we discuss how the properties of SESN progenitors can be constrained through different phases of the bolometric light curve. We find that the light curve rise time is strongly impacted by the strength of radioactive nickel mixing and treatment of helium recombination. These can vary between events and are often not accounted for in simpler modeling approaches, leading to large uncertainties in ejecta masses inferred from the rise. Motivated by this, we focus on the late time slope, which is determined by gamma-ray leakage. We calibrate the relationship between ejecta mass, explosion energy, and gamma-ray escape time $T_0$ using a suite of numerical models. Application of the fitting function we provide to bolometric light curves of SESNe should result in ejecta masses with approximately 20\% uncertainty. With large samples of SESNe coming from current and upcoming surveys, our methods can be utilized to better understand the diversity and origin of the progenitor stars.

    
\end{abstract}


\section{Introduction}
\label{sec:intro}

The broad classification of stripped-envelope supernovae (SESN) refers to core-collapse supernovae (CCSNe) that have lost most or all of their outer hydrogen (SN~IIb and SN~Ib, respectively) and possibly helium (SN~Ic) envelopes prior to explosion \citep[for reviews of SN classification, see][]{Filippenko1997,Gal-Yam2017}. Understanding the progenitors of these events and the ways in which they differ from more traditional SNe~II provides insight into the life cycles of massive stars and what events impact their evolution near death. In a few instances, Type~Ibc progenitors have been identified in pre-explosion imaging \citep[e.g.,][]{Eldridge2016, VD2018, Kilpatrick2018, Xiang2019}, but in the vast majority of cases the progenitors are too dim and distant to use such techniques. It is therefore crucial to use the SNe themselves to study the properties of their progenitors. 

One especially important parameter is the ejecta mass, $M_{\rm ej}$, because it can be used to discriminate between different origins of SESNe and relating it to the zero age main sequence (ZAMS) mass, $M_{\rm ZAMS}$, teaches us about the mass loss process. The ZAMS mass ranges covered by different classes of SESN is still unclear, as well as whether or not those mass ranges represent discrete groups of progenitors or a continuous distribution \citep{Ouchi2017}. A continuous distribution of ZAMS masses between SESN subtypes may indicate common progenitors that undergo various amounts of stripping through binary evolution, rather than single-star evolution \citep{Lyman2016}. Single-star evolution requires a radiatively driven wind to remove the outer envelopes, which is only possible in especially massive stars, however the number of observed SESN is too high to explain the majority with such high mass stars \citep{smith2011, smith2014, sravan2019, Dessart2020}, and that traditional stellar wind prescriptions cannot readily explain high mass loss rates inferred for certain progenitors \citep{Ouchi2017}. Newer results from \citet{Fang2019}, which are further supported by \citet{Ning-Chen2022}, suggest a hybrid process to generate SN~Ic progenitors where the hydrogen envelope is stripped via binary interactions and the helium envelope is driven off by winds. This would still require sufficiently massive stars, though the exact ZAMS masses needed are still debated. Searching for any correlations between explosion parameters and SESN classes brings us closer to understanding these progenitors and the late stages of evolution for massive stars.

Two of the most common ways to constrain $M_{\rm ej}$ from SN light curves is using the rise time \citep[e.g.,][]{arnett1982} and the decaying tail \citep[e.g.,][]{Colgate1980a}. As we further argue below in Section~\ref{sec:rise}, using the rise can be less reliable due to differences between events that are not captured in many simple fitting models. \cite{wheeler2015} shows empirically that there are inconsistencies in the parameters derived from the rise and tail of SESNe. Motivated by this, we focus on the decaying tail method and calibrate it to numerical models so that it can be better utilized to estimate $M_{\rm ej}$. In Section~\ref{sec:models}, we describe the suite of numerical SESN models we ran and our methods for constraining the relationship between $M_{\rm ej}$, explosion energy, and late time slope. In Section~\ref{sec:results}, we report and discuss our results, and in Section~\ref{sec:params}, we explore the uncertainties in the derived $M_{\rm ej}$ introduced in our numerical models. In Section~\ref{sec:2011dh} we apply our method to the event 2011dh, a well studied SN~IIb with broadband late-time data and compare our results to those presented in \cite{Ergon2015}. In Section~\ref{sec:conclusion} we conclude with a summary of our results.

\section{Rise vs Decline}
\label{sec:rise}


As described above, the most common ways to constrain $M_{\rm ej}$ from SN light curves are via the rise and tail. It has been noted across many studies that it is difficult to match parameters derived from the rise and decline phases of CCSNe light curves \citep[see][and others]{Ensman1988,  Arnett1989,  Woosley1994, Woosley2021, Clocchiatti1997, wheeler2015, Fang2019}. Here we compare and contrast how different physical assumptions can impact the $M_{\rm ej}$ inferred from both of these methods.

The often utilized relationship between ejecta mass and rise time of the light curve is \citep{arnett1982}:

\be
    t_{\rm rise} \propto \lp \frac{\kappa M_{\rm ej}}{v_{\rm ph}c} \rp ^{1/2} \label{eq:t_rise},
\ee
where $\kappa$ is the opacity, $v_{\rm ph}$ is the velocity of the photosphere, and $c$ is the speed of light. To understand how various modeling assumptions impact the shape of the early light curve, we compare with a fiducial model of a $12 M_{\odot}$ ZAMS star that experienced mass transfer with a binary companion in the open-source code \texttt{SNEC} \citep{SNEC}. The core of the star, defined as the mass coordinate of the Si/O shell in a given model, is excised to represent the formation of a neutron star. The nickel mass is set to be 0.1$M_{\odot}$ and is mixed out to a mass coordinate of 0.5$M_{\rm ej}$. We choose this boundary to represent moderate mixing. More details on our numerical setup are provided in Section~\ref{sec:models}. We next vary the properties of the explosion around this fiducial model to explore the impact on the rise and tail.

\subsection{Recombination}
\label{sec:recomb}

\begin{figure}
\includegraphics[width=0.48\textwidth]{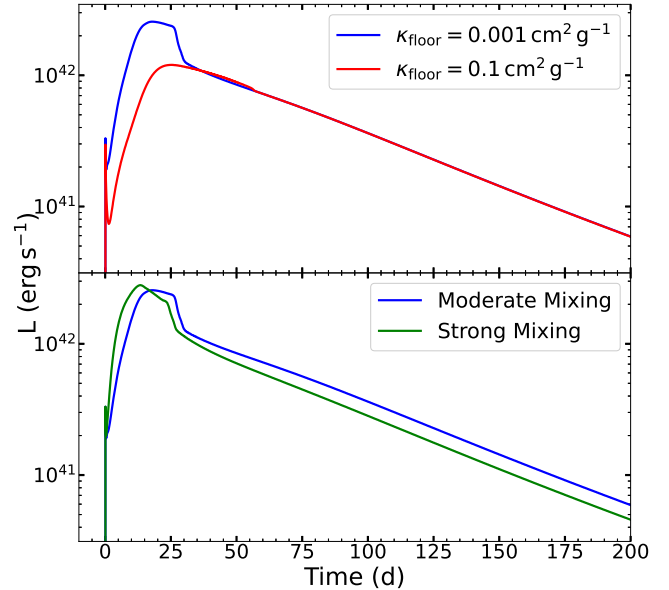}
\caption{(Top) Two simulations of our $M_{\rm ZAMS} = 14M_{\odot}$ model in which we vary the opacity floor, effectively controlling the amount of recombination that is allowed to happen in the outer layers of the ejecta. The blue light curve, which represents significant recombination, rises significantly faster and to a brighter peak than the red (no recombination) light curve, but the two begin declining together at ~60 days . (Bottom) Two simulations of our $M_{\rm ZAMS} = 14M_{\odot}$ model in which we vary the strength of $^{56}{\rm Ni}$ mixing. As expected, the stronger mixing light curve (green) rises faster and to a slightly brighter peak. While the two light curves do not totally decline together in the radioactive decay phase, the tails have nearly the same slope.}
\label{fig:recomb/mix}
\end{figure}

 Many semi-analytic modeling techniques assume a constant opacity everywhere within the ejecta, directly impacting the ejecta mass derived for a given rise time. This assumption can be limiting because opacity evolves both spatially and in time with the temperature and composition of the ejecta, which may be different between the SESN subtypes \citep{Lyman2016}. \cite{wheeler2015} argues that recombination near light curve peak can decrease the opacity in outer, recombined layers and impact the relationship between $t_{\rm rise}$ and $M_{\rm ej}$ (see also \citealp{Piro2014}). We explore the impact of recombination by varying the minimum opacity value we allow \texttt{SNEC} to reach, effectively allowing more or less recombination to occur in the outer layers of the ejecta. The effect on the light curve rise is seen in the upper panel of Figure~\ref{fig:recomb/mix}, where we compare a low opacity floor of 0.001$\,{\rm cm}^2\,{g}^{-1}$ to a larger opacity floor of 0.1$\,{\rm cm}^2\,{g}^{-1}$ (roughly the value of singly ionized helium-rich material). This shows that for otherwise identical models, an increased opacity floor causes the the rise time to be slower. Despite this, the decay tails of the two models coincide and are not affected by whether recombination is included in the explosion model. 

Another way to think about this is that the rise time tracks the mass contributing to diffusion rather than the total ejecta mass. One implication from \cite{Lyman2016} is that SNe~Ib and SNe~Ic have similar distributions of explosion parameters, in particular $M_{\rm ej}$, which may indicate a common progenitor. However, because helium has a higher ionization temperature than carbon and oxygen, for a given temperature evolution, models with a significant helium envelope leftover (SNe~Ib) are more likely to experience recombination, causing a smaller amount of mass to contribute to diffusion. With this in mind, it is possible that the results of \cite{Lyman2016} actually indicate that SNe~Ib and SNe~Ic have similar diffusion masses, while there is a portion of the ejecta mass above the photosphere with $\kappa \sim 0$ that is not being probed in SNe~Ib.

\begin{figure*}
\includegraphics[width=0.99\textwidth,trim=0.0cm 0.0cm 0.5cm 0.0cm, clip]{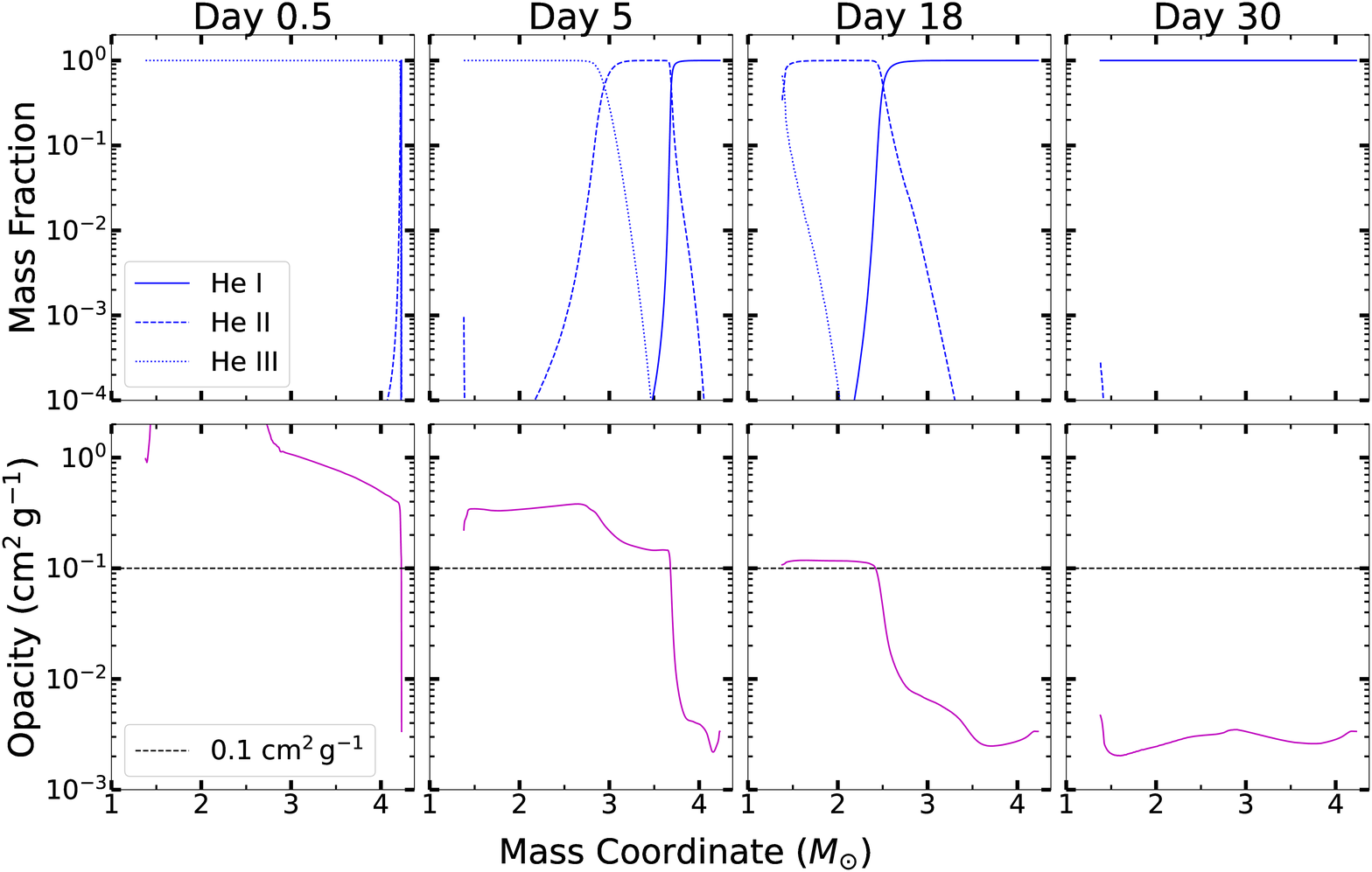}
\caption{Snapshots of the ionization (top row) and opacity (bottow row) profiles of our fiducial model at 0.5, 5, 18, and 30 days beyond the explosion epoch. The bolometric luminosity of this model peaks at roughly 18 days, by which time a significant amount of the outer, helium-rich ejecta has recombined into neutral helium and the opacity continues to decrease. By day 30 the ejecta is entirely recombined and the opacity profile has flattened out.}
\label{fig:ionization}
\end{figure*}

Figure~\ref{fig:ionization} shows the progression of the recombination wave as it moves inward through the expanding and cooling ejecta. As the helium-rich ejecta recombines the opacity profile drops off sharply, which causes the rise time to decrease according to Equation~(\ref{eq:t_rise}) as photons are able to diffuse out more quickly through the transparent material. By the time this model reaches its peak bolometric luminosity at roughly 18 days, a significant amount of the ejecta has recombined and the outer opacity profile has fallen well below $\mathbf{0.1\,{\rm cm}^2{\rm g}^{-1}}$, roughly the electron scattering value for singly ionized helium.

\subsection{Nickel Mixing}
\label{sec:mixing}

The strength of mixing of $^{56}{\rm Ni}$ and how to quantify it is not well agreed upon in CCSNe modeling, with some methods restricting mixing to most central regions of ejecta to better match results of 3D numerical modeling \citep[which we will refer to as weak mixing, such as in ][]{arnett1982,valenti2008,Lyman2016,medler2021}, while others allow mixing to the outermost edges of the ejecta (which we will refer to as strong mixing, such as in \citealp{Dessart2020}). It is therefore difficult to compare models across the literature due to the strong effect that mixing has on the early light curve and therefore on the ejecta mass estimate in constant-opacity models. Nevertheless, the choice of how $^{56}{\rm Ni}$ mixing is treated has a strong impact on the early light curve because it determines how efficiently the gamma-rays from its decay heats the ejecta. In strongly mixed models, the heating wave reaches the outer envelope faster, resulting in a faster rise to the light curve peak and dimmer decay tail \citep{meza2020}. This also increases the likelihood of gamma-ray leakage as there is less ejecta to trap gamma-rays near the surface. 

In the lower panel of Figure~\ref{fig:recomb/mix}, we compare otherwise identical SESN models with varying mixing strengths. We control this in our simulations by defining a boundary in mass space out to which the total nickel mass (also set in our simulation) is distributed prior to our compositional smoothing procedure, which uses a ``boxcar'' approach with a boxcar width of $0.4M_{\odot}$ over 4 iterations (the same parameters defined in \citealp{SNEC}). Models with weaker mixing, corresponding to a boundary well inside of the ejecta, have a resulting nickel distribution that is more heavily concentrated at the inner ejecta boundary, whereas models with stronger mixing have a flatter distribution across the ejecta. 

Models that are mixed out to half of their ejecta (which we define as moderately mixed) rise more slowly than the strongly mixed models. One would therefore infer a larger mass for the moderately mixed model when utilizing Equation (\ref{eq:t_rise}). The light curves do, however, have similar slopes at late times, though at a different normalization. This indicates that estimates of ejecta mass from the late time light curve may be robust against assumptions related to mixing strength, however provide a poorer estimate of the overall nickel mass. This concept is discussed further in Section~\ref{sec:params}. In the models utilized going forward, we use a moderate mixing prescription of allowing $^{56}{\rm Ni}$ out to $50\%$ of the ejecta mass. This gives us a baseline to which we can compare other mixing methods across the literature.



\subsection{Decay Tail}
\label{sec:tail}

Given the shortcomings of utilizing the light curve rise as discussed above, we instead focus on the late light curve slope. Similar procedures have been applied to analyzing thermonuclear events (SNe~Ia) as a method for constraining progenitor models and explosion mechanisms \citep[see][]{Colgate1980a, Cappellaro1997, Clocchiatti1997, Jeffery1999, Milne1999, Milne2001, Stritzinger2006}. Here we quickly summarize how gamma-ray leakage impacts the late time slope using the analytic formalism presented in \cite{wheeler2015}. The radioactive heating rate is given by

\be
    L_{\rm Ni} (t) = \frac{M_{\rm Ni}}{M_{\odot}}\left[ \epsilon_{\rm Ni} e^{-t / \tau_{\rm Ni}}+\epsilon_{\rm Co} \lp e^{-t/\tau_{\rm Co}} - e^{-t / \tau_{\rm Ni}}\rp \right], \label{eq:Lheat}
\ee
where $\epsilon_{\rm Ni} = 3.97 \times 10^{10}\,{\rm erg}\,{\rm g}^{-1}$ and $\epsilon_{\rm Co} = 7.29\times 10^{9}\,{\rm erg}\,{\rm g}^{-1}$ are the decay energies produced in one second by one gram of $^{56}{\rm Ni}$ and $^{56}{\rm Co}$ respectively, and $\mathbf{\tau_{\rm Ni} = 8.8\,}$days and $\tau_{\rm Co} = 111\,$days are their decay times. 


The strength of gamma-ray leakage is controlled by the gamma-ray diffusion timescale \citep{Clocchiatti1997} given by
\be
    T_0 = \lp \frac{\eta \kappa_\gamma M_{\rm ej}^2}{E_k} \rp^{1/2}, \label{eq:T0_1}
\ee
where $\kappa_{\gamma} = 0.06 Y_e {\rm cm}^2\,{\rm g}^{-1}$ is the gamma-ray opacity, $Y_e$ is the electron fraction, and $\eta$\footnote{In \cite{wheeler2015} this constant is referred to as C and is equal to 0.05, which comes from the assumption of a power-law density distribution. In Section~\ref{sec:models} we will introduce our own parameter C and therefore use $\eta$ here to avoid confusion. In the following sections we will solve for a new value of $\eta$ using our numerical simulations.} is a constant that depends on the density profile. A key point is that $\kappa_{\gamma}$ is largely independent of the density and temperature of the ejecta. This helps make the gamma-ray leakage a better probe of the total ejecta mass.


It is useful to rewrite Equation~(\ref{eq:T0_1}) in terms of parameters that are more easily observable. Utilizing that for a constant density sphere $E = (3/10) M_{\rm ej} v_{\rm ph}^2$ \citep{arnett1982}, where $v_{\rm ph}$ is the velocity of the photosphere, we find 
\be 
T_0 = \lp\frac{10 \eta \kappa_{\gamma} M_{\rm ej}}{3 v_{\rm ph}^2}\rp^{1/2}. \label{eq:T0_2}
\ee
The energy deposition rate of gamma-rays that are able to thermalize is therefore
\be
    L_{\rm heat} = L_{\rm Ni} [ 1 - e^{-(T_0/t)^2} ]. \label{eq:Ldecay}
\ee
Therefore $T_0$ controls the slope of the late-time light curve where smaller values of $T_0$ corresponds to a short leakage time such that gamma-rays are less trapped within the ejecta, causing a steeper decline in the decay tail. Conversely, as $T_0 \rightarrow \infty$ (corresponding to no gamma-ray leakage), the portion of Equation~(\ref{eq:Ldecay}) in brackets asymptotes to 1. We concede that, as is explored in \cite{Sharon2020}, Equation~\ref{eq:Ldecay} is not a sufficient description of gamma-ray deposition at all times. However, we show in the following analysis that in the late time regime where the light curve is dominated by the decay of $^{56}$Co, Equation~\ref{eq:Ldecay} is appropriate for fitting $T_0$.

We emphasize that, while the late time light curve is normalized by the total nickel mass, the slope of the decay tail is independent of it and the degree to which it is mixed. We can show this by summarizing the arguments of \cite{Clocchiatti1997}, which analyze the derivative of $\ln({\rm L_{\rm heat}})$ with respect to time at late times. In this limit, where $e^{-t/\tau_{\rm Ni}} \ll e^{-t/\tau_{\rm Co}}$, we see that

\be
    \begin{gathered}
        \frac{d \ln(L_{\rm heat})}{dt}  \approx \frac{1}{L_{\rm heat}} \frac{d}{dt}[M_{\rm Ni} \epsilon_{\rm Co} e^{-t/\tau_{\rm Co}}(1 - e^{-(T_0/t)^2})] \\
         \approx \frac{-1}{\tau_{\rm Co}} - \frac{2 T_0^2}{t^3}\frac{e^{-(T_0/t)^2}}{1 - e^{-(T_0/t)^2}}. \label{eq:slope}    
    \end{gathered}
\ee 

This indicates that the logarithmic slope is independent of the total nickel mass and that incomplete gamma-ray trapping results in a steeper slope than just simple cobalt decay. This slope is not constant but instead is time dependent, as shown by the term on the right side of Equation~(\ref{eq:slope}). Nevertheless, we can identify when the slope is most slowly varying by taking the derivative of the right term in Equation~(\ref{eq:slope}) with respect to time and setting it to zero. Performing this numerically, we indeed find that this occurs at $t\approx1.07 T_0$, in agreement with \cite{Clocchiatti1997}. When t is close to $T_0$, we expect the slope to be slowly varying and the gamma-ray trapping impact to be a constant factor that steepens the slope. Evaluating Equation~(\ref{eq:slope}) at $t=T_0$, 
\be
    \frac{d \ln(L_{\rm heat})}{dt} \approx \frac{1}{\tau_{\rm Co}} - \frac{-1.17}{T_0}.
\ee
Typical values of $T_0$ are $\sim$100 \textemdash 150 days \citep{wheeler2015}, though for lower mass progenitors or higher energy explosions, $T_0$ values are smaller and the quality of this approximate begins to degrade. Still this analysis motivates that we should fit late time SN light curves to get the most robust constraint on $T_0$. We discuss our exact fitting procedure in more detail in the following section. 

The above discussion, in unison with the many works cited throughout this section, signify that fitting the light curve tail has advantages for constraining the ejecta mass rather than using the light curve rise. To quickly summarize, we find the rise depends sensitively on the distribution of nickel and the treatment of recombination, which is often absent from simple semi-analytic models. This is not even accounting for non-gray opacity, and the opacity of UV/optical/IR photons can be very complicated and frequency dependent during the rise. In contrast, the gamma-ray opacity is largely independent of the thermal state of the ejecta, and we find agreement with \cite{Clocchiatti1997} that around $t\approx T_0$, incomplete gamma-ray trapping results in a nearly constant steepening of the late time slope. Given the relationship between $T_0$ and $M_{\rm ej}$ summarized in Equation~(\ref{eq:T0_2}), fitting for $T_0$ should allow for a strong constraint on the ejecta mass. However, to do this effectively, we must better understand the relationship between $T_0$, $M_{\rm ej}$, and $v_{\rm ph}$ in more detail than what is simply given in Equation~\ref{eq:T0_2}, which we explore next. 

\section{Models and Methods}
\label{sec:models}

The framework described in Section~\ref{sec:rise} demonstrates how fitting for $T_0$ can be used to measure the ejecta mass of SESNe, but although Equation (\ref{eq:T0_2}) provides a simple relationship between $T_0$, $M_{\rm ej}$, and $v_{\rm ph}$, in detail we expect this relationship to be more complicated for realistic models. Motivated by this, we next calibrate this relationship using numerical simulations. Motivated by the power-law dependence in Equation~\ref{eq:T0_2}, we begin with the parameterization
\be
    \log T_0 = A + B\log M_{\rm ej} + C\log v_{\rm ph}, \label{eq:log}
\ee
where $A$, $B$, and $C$ are free parameters. This functional form captures the power law-like dependence expected from analytic arguments. If we were to derive these parameters from Equation~(\ref{eq:T0_2}), we would have $A\approx -1.151$, $B= 1/2$, and $C=-1$. We will compare these values to those derived in our work in Section~\ref{sec:results}. 

To improve on these analytic results, we calibrate Equation~(\ref{eq:log}) to numerical simulations. We explode a grid of SESN models using \texttt{SNEC} \citep[the SuperNova Explosion Code,][]{SNEC} at various energies to produce 77 light curves that may be fit for $T_0$. We utilize the pre-explosion models from the suite of binary evolution models published in \cite{Laplace2021}. These models cover the mass range $M_{\rm ZAMS} = 11-21\, M_{\odot}$ at intervals of roughly $1 M_{\odot}$, and were evolved using the stellar evolution code \texttt{MESA} \citep{MESA}, employing a fully coupled nuclear network of 128 isotopes beyond the point of core oxygen burning. The models underwent stable, case B mass transfer with a companion star of $80\%$ the initial mass of the primary, in orbital periods between 25 \textemdash 30 days. See \cite{Laplace2021} for more details on modeling conditions.

\begin{figure}
\includegraphics[width=0.46\textwidth]{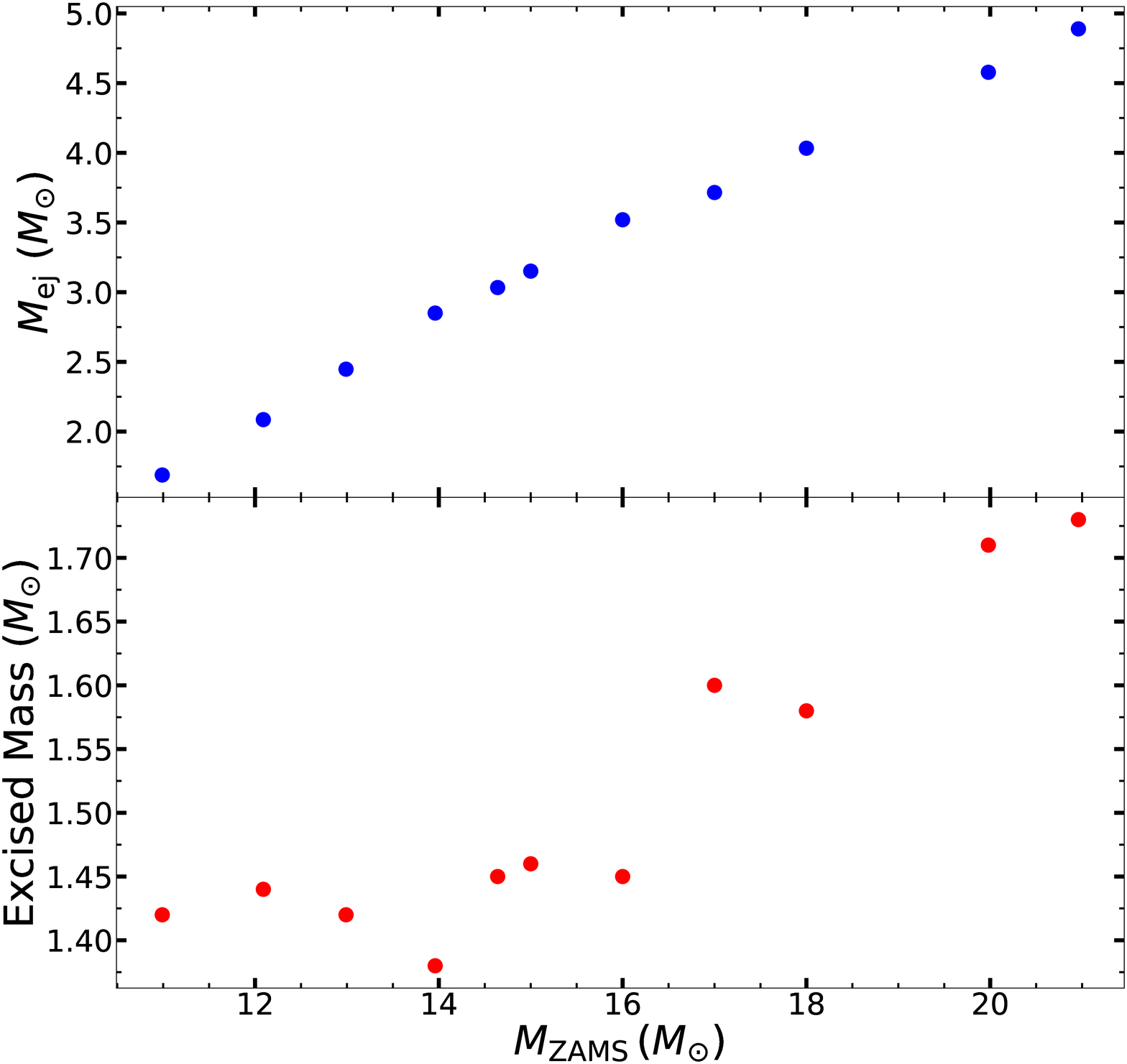}
\caption{(Top) The ejecta masses of each model at the onset of core-collapse compared to their ZAMS mass, where ejecta mass refers to the mass of the model after completing mass transfer with its companion and iron core excision. (Bottom) The mass of the iron core excised from each model compared to ZAMS mass.}
\label{fig:masses}
\end{figure}

 Following the prescription of \cite{measuring}, we excise the core from the pre-explosion model to form a neutron star at the Fe/Si boundary, and smooth the remaining compositional profile with a ``boxcar'' approach using the same parameters as in \cite{SNEC}. Regarding mixing strength, the nickel mass boundary (previously defined as $0.5 M_{\rm ej}$ for moderate mixing and $M_{\rm ej}$ for strong mixing) sets the mass coordinate out to which the nickel is distributed before it is smoothed by the boxcar procedure.  Figure~\ref{fig:masses} shows the ejecta mass of each model after core excision compared to ZAMS mass (top panel) and the mass of the excised core compared to ZAMS mass (bottom panel). The resulting ejecta masses are in the range of $2-5 {\rm M_{\odot}}$, which roughly matches what is inferred for the majority of Type~Ibc \citep{Lyman2016}.
 
 Our adopted opacity floors and the mixing of $^{56}{\rm Ni}$ are discussed in Sections~\ref{sec:recomb} and ~\ref{sec:mixing} respectively. Finally, we use a ``thermal bomb'' mechanism with an energy injection of $1.0 - 4.0 \times 10^{51}\,{\rm erg}$ at intervals of $5 \times 10^{50}\,{\rm erg}$, the range of which is inferred from the explosion parameters presented in \cite{Lyman2016}. We modeled each supernova for 225 days, well into the cobalt dominated phase. 

\begin{figure}
\includegraphics[width=0.46\textwidth]{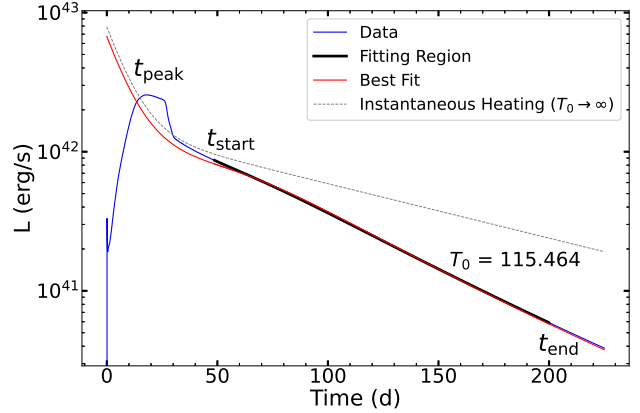}
\caption{Here we demonstrate our fitting procedure on our $M_{\rm ZAMS}=14 M_{\odot}$ fiducial model, shown in blue. The black line shows the time over which we are fitting for $T_0$ and the red curve shows Equation~(\ref{eq:Ldecay}) calculated with the best fit value of $T_0$ = 115.464.}
\label{fig:fit}
\end{figure}

Our fitting procedure is as follows (also see Figure~\ref{fig:fit} for a schematic demonstrating our methods). As motivated by the discussion in Section~\ref{sec:tail}, we need to perform the fit around the time $t\approx T_0$. Since we do not know $T_0$ a priori, we want to make sure that we are well past the SN peak and in the tail. We do this by choosing to begin the $T_0$ fitting at roughly half the time it takes for the ejecta to become thin to optical photons

This is convenient because of the relationship between optical thinness and the light curve will peak. The rise time expressed in Equation~\ref{eq:t_rise} is proportional to the diffusion time and describes when the light curve will reach its peak. Using that $\tau \approx \kappa \rho r \approx \kappa M /r^2$ and that $r \approx vt$, we can solve for the time that the ejecta becomes optically thin by allowing $\tau \rightarrow 1$. We see then that

\be
    t_{\rm thin} \approx \frac{(\kappa M)^{1/2}}{v} \approx t_{\rm rise}(c/v)^{1/2}
\ee
where we approximate $c/v \approx 30$. We choose $t_{\rm start} = 0.5 t_{\rm thin}$ to ensure that the fit begins far enough beyond the time of peak to only sample the decline and because it should be simple to apply to any set of data that constrains the time of explosion. 

We chose to end our fitting at $t_{\rm stop} = 200\ {\rm days}$ beyond the time of explosion. While this means that models that peak at different times are fit over a slightly different number of days, we find that the exact length of time that is fit is less important than fitting out to sufficiently late time. In Section~\ref{sec:params}, we discuss the consequences of changing $t_{\rm stop}$. 

Finally, in the following analysis, we consider the characteristic photosphere velocity for each model at the time of the light curve peak. We use this velocity, rather than that at $t_{\rm start}$, because by this time in our lower mass models, the photosphere has already fallen down to the inner mass boundary of the star. \cite{pm2014} demonstrate that, for a sample of SNe~IIb and SNe~Ib, the velocity at the thermalizaion depth (also referred to as color depth) is similar to, but less than the He I velocity. Since the color depth is slightly deeper than the photosphere, they conclude that the helium is tracking the photosphere in these H-poor events. This may be utilized for events observed in the future.

\section{Fitting}
\label{sec:results}

\begin{figure}
\includegraphics[width=0.48\textwidth]{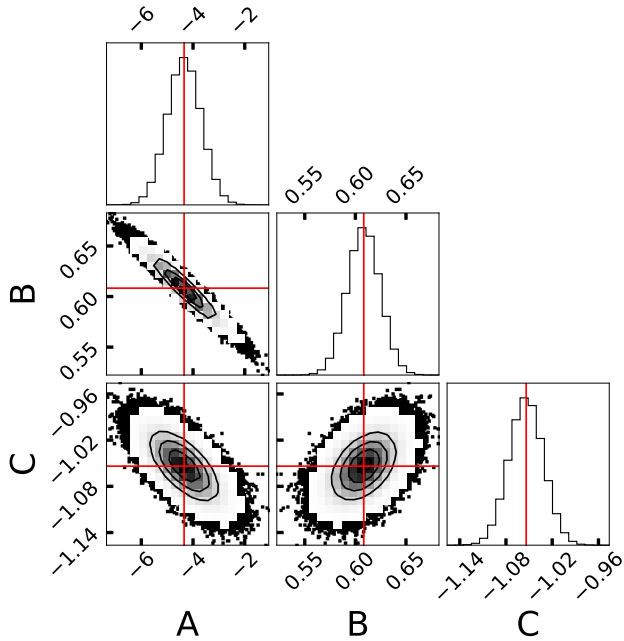}
\caption{Correlations between the values of $A$, $B$, and $C$ show that the value of $A$ is heavily dependent on the values of both $B$ and $C$, while $B$ and $C$ depend little on each other. This contributes to the large confidence interval around our best fit value of $A$.}
\label{fig:corner}
\end{figure}

We fit all 77 models in our suite of explosion simulations to Equation (\ref{eq:Ldecay}). This is done by fitting for the best values of $M_{\rm Ni}$ and $T_0$ for each simulated light curve, independent of whether this is the true $M_{\rm Ni}$ value (to replicate how one would treat fitting an observed light curve). We solve for the values of $A$, $B$, and $C$ that best allow our grid of fits to match Equation~(\ref{eq:log}) by using both the python optimization function \texttt{scipy.optimize} and the Markov Chain Monte Carlo (MCMC) python package \texttt{emcee} to approximate the posterior probability density function of each parameter. The optimized values are found to be
\be
    \begin{aligned}
        A = -4.34^{+0.65}_{-0.65} \\
        B = 0.608^{+0.02}_{-0.02} \\
        C = -1.05^{+0.02}_{-0.02}. \label{eq:opt}
    \end{aligned}
\ee
These are roughly consistent with the values of $B$ and $C$ expected from simple analytic arguments in Section~\ref{sec:models}, but slightly different due to the more realistic density and velocity profiles captured by our simulations. Furthermore, the value of $A$ is significantly smaller than what is commonly used in the literature. Figure~\ref{fig:corner} shows 2D projections of the parameter space explored by the MCMC chain and demonstrates the relationships between each parameter. The Pearson's correlation coefficients of the sample are $\rho_{\rm A,B} = -0.962$, $\rho_{\rm A,C} = -0.663$,  and $\rho_{\rm B,C} = 0.434$. 

Using these values of $A$, $B$, and $C$, we can now write an updated version of Equation~(\ref{eq:T0_2}):
 \be
    T_0 = \lp\frac{10}{3}\rp ^{1/2} \Tilde{\eta} \kappa_\gamma M_{\rm ej}^{0.608} v_{\rm ph}^{-1.05},\label{eq:T0_3}
 \ee
where we now have $\Tilde{\eta} \approx  4.51 \times 10^{-4}$, which is 2 order of magnitude smaller than the commonly used value. 
 
\begin{figure}
\includegraphics[width=0.48\textwidth]{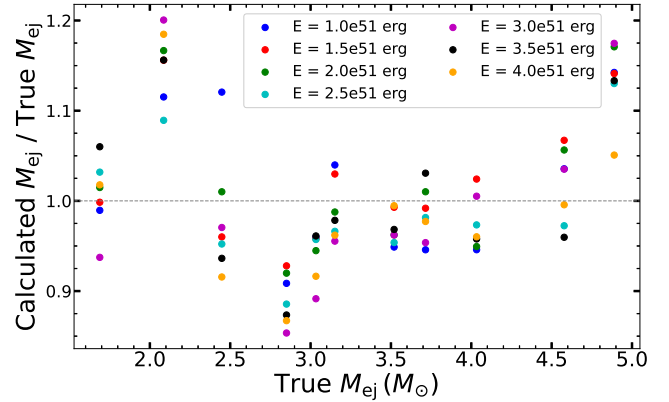}
\caption{Ratio of mass calculated from Equation~(\ref{eq:T0_3}) utilizing our optimized values of $A$, $B$, and $C$ in Equation~(\ref{eq:opt}). Calculated ejecta masses are within 20\% of the true mass across our full grid of models.}
\label{fig:results_mean}
\end{figure}

To test the ability of this technique to constrain $M_{\rm ej}$, in Figure~\ref{fig:results_mean} we plot the ratio of $M_{\rm ej}$ inferred from Equation~(\ref{eq:T0_3}) to the true $M_{\rm ej}$. This demonstrates that the calculated masses are mostly within $10\%$ of the true masses with some outliers at the $\sim 15$ \textemdash 20\% level. Because the range of possible ejecta masses for SESN progenitors is so large, the uncertainties related to Equation~(\ref{eq:T0_3}) are small enough to still provide useful constraints.

\section{Checking the Robustness of this Technique}
\label{sec:params}

As discussed in Section~\ref{sec:rise}, fitting the late time slope should be more robust for inferring $M_{\rm ej}$ than using the early rise. In practice though, uncertainties in the underlying SESN model and practical issues with fitting may result in less reliable $M_{\rm ej}$ constraints. Next we test how uncertainties in the amount of mixing of $^{56}{\rm Ni}$ impacts the inferred $M_{\rm ej}$. We also repeat our fitting procedure for $t_{\rm stop} = 100$ days for cases where light curve photometry cannot be attained for the full 200 days as we did above. 

\begin{figure}
\includegraphics[width=0.47\textwidth,trim=0.0cm 0.0cm 0.5cm 0.0cm, clip]{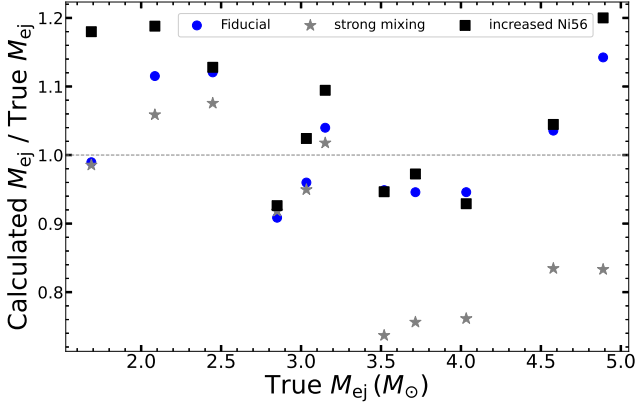}
\caption{Ratio of ejecta mass calculated from Equation~(\ref{eq:log}) utilizing our best fit values of $A$, $B$, and $C$ for cases varying the $^{56}{\rm Ni}$ mixing and concentration. We see that increasing the mixing strength tends to result in a slightly lower ejecta mass estimate, but is still within a reasonable error.}
\label{fig:other_params}
\end{figure}

We address questions relating to $^{56}{\rm Ni}$ by running additional sets of explosion simulations. In the first, we allow the  $^{56}{\rm Ni}$ to mix evenly throughout the entire ejecta. In the second, we double the amount of  $^{56}{\rm Ni}$ to 0.2$M_{\odot}$ and keep it mixed within the inner half of the ejecta as before. Each set is run for the same set of values of $M_{\rm ZAMS}$ and with an explosion energy of $10^{51}$ erg. Our results from fitting these additional simulations with Equation~(\ref{eq:log}) and using Equation~(\ref{eq:T0_3}) for $M_{\rm ej}$ are summarized in Figure~\ref{fig:other_params}, along with a set of previous simulations using the same explosion energy. This shows that the calculated $M_{\rm ej}$ are still within the range of 20\% of the true $M_{\rm ej}$. Increasing the $M_{\rm Ni}$ tends to increase the inferred $M_{\rm ej}$, while stronger mixing tends ot result in characteristically lower $M_{\rm ej}$ estimates. 

We emphasize that when performing the fitting of the late time slope, we keep the normalization as a free parameter. This means we are effectively fitting an $M_{\rm Ni}$ value, independent of what the true $M_{\rm Ni}$ is. In fact, the fit $M_{\rm Ni}$ is always lower than the $M_{\rm Ni}$ present in a given simulation because mixing puts nickel above the photosphere and unable to heat the ejecta. In principle, one could attempt to fit the slope and also $M_{\rm Ni}$ using the integrated bolometric light curves (e.g.,\citealp{Khatami2019}). Since our main objective here is understanding how well we can constrain the ejecta mass, we forego attempting to also fit the total nickel mass in detail.

\begin{figure}
\includegraphics[width=0.47\textwidth,trim=0.0cm 0.0cm 0.5cm 0.0cm, clip]{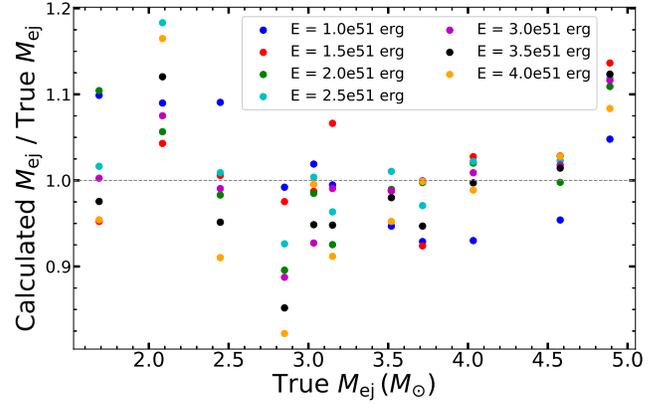}
\caption{Ratio of calculated to true ejecta masses using Equation~(\ref{eq:log}) with $A$, $B$, and $C$ values in Equation~(\ref{eq:100}) for models that are fit over the time frame $0.5 t_{\rm thin}$ to 100 days beyond the time of explosion. Similar to the 200 day fit, these results recover $M_{\rm ej}$ values to within 20\%.} 
\label{fig:short}
\end{figure}

\begin{figure}
\includegraphics[width=0.47\textwidth,trim=0.0cm 0.0cm 0.5cm 0.0cm, clip]{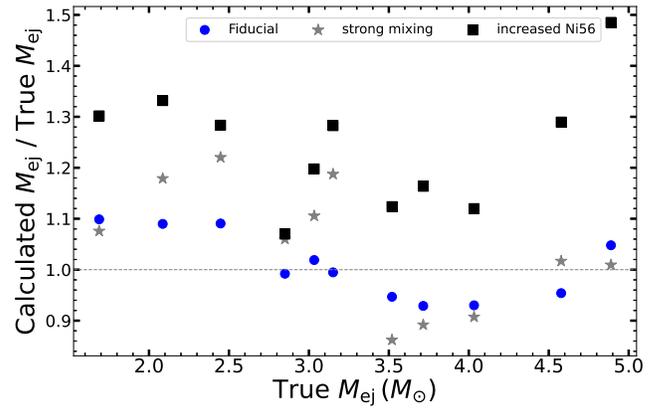}
\caption{Ratio of calculated to true ejecta masses using Equation~(\ref{eq:log}) with $A$, $B$, and $C$ values in Equation~(\ref{eq:100}) for models varying the $^{56}{\rm Ni}$ mixing and concentration. These values perform worse than the 200 day fit with varying the nickel parameters, particularly those of high mass.}
\label{fig:other_short}
\end{figure}

Constructing the late time bolometric light curves of SESNe can be challenging, especially at the $t_{\rm stop} = 200$ days we use for our fitting procedure. To test the impact of a smaller fitting window, we repeat our analysis covered in Section~\ref{sec:models} using $t_{\rm stop} = 100$ days instead. We find best fit values
\be
    \begin{aligned}
        A_{100} = -4.01^{+0.65}_{-0.65} \\
        B_{100} = 0.575^{+0.02}_{-0.02} \\
        C_{100} = -0.967^{+0.02}_{-0.02}. \label{eq:100}
    \end{aligned}
\ee These values are also highly correlated and return estimates of $M_{\rm ej}$ within a similar margin of error, as seen in Figure~\ref{fig:short}. However, this shorter fitting procedure breaks down for models where we vary the $^{56}{\rm Ni}$ mixing and concentration, particularly those of higher mass, as seen in Figure~\ref{fig:other_short}, where the error in the calculated $M_{\rm ej}$ increases significantly. This may be because higher mass models peak much later than less massive models and therefore stopping fitting at 100 days is too short to sample the late time behavior.

Previously, \cite{Taddia2018} utilized the \cite{wheeler2015} method to analyze SESNe in the Carnegie Supernova Project sample (CSP-1), but used a fitting time frame that ended sometimes as short as 60 days past the explosion epoch. Our findings indicate that this time frame, especially for more massive events, is not far enough beyond the light curve peak to sufficiently fit the late time slope and is susceptible to influence from the $^{56}{\rm Ni}$ parameters that are not considered in Arnett-like models. 

Finally, to understand how universally applicable each set of $A$, $B$, and $C$ values are, we calculated ejecta masses for each of the fitting time frames using the opposite set of best fit parameters. We find that the 200 day fit of ($A, B, C$) (values in Equation~(\ref{eq:opt})) can be applied to the models fit to only 100 days and still return estimates of ejecta masses within $20\%$ error, while the $A$, $B$, and $C$ values from the 100 day fit (Equation~(\ref{eq:100})) do not perform well when applied to the 200 day data. This indicates that the values we have provided in Equation~(\ref{eq:opt}) may be widely applied to data sets that track events for less than 200 days, but again we emphasize that later time data is important for appropriately accounting for nickel mixing. 

\section{Comparison to SN 2011dh}
\label{sec:2011dh}

SN 2011dh is a SN~IIb that exploded on May 31, 2011 in the nearby galaxy M51 \citep{Arcavi2011}. Its light curve out to nearly 100 days has been extensively studied in \cite{Bersten2012} and \cite{Ergon2014} using both hydrodynamical and spectral modeling. They conclude that SN 2011dh had a helium core mass of $M_{\rm He} = 3.31_{-0.57}^{+0.54} M_{\odot}$ and exploded with a kinetic energy of $E=6.4_{-3.0}^{+3.8} \times 10^{50}$ erg and a nickel mass of $M_{\rm Ni} = 0.075_{-0.020}^{+0.028} M_{\odot}$, which was very strongly mixed out through 95\% of the ejecta. The yellow supergiant progenitor is estimated to have a ZAMS mass of 12 \textemdash 15 $M_{\odot}$. Photometric and spectral follow up of this event is presented in \cite{Ergon2015}, which provides well-sampled, broadband data beyond 400 days past the explosion epoch, making this event a good target for our method. 

The bolometric light curve of 2011dh is derived from a combination of a spectroscopic method and a photometric method, where the photometric method is utilized at times or in wavelength ranges where spectroscopic data is unavailable (see \citealp{Ergon2014, Ergon2015} for details). Previous studies define the photospheric radius as the radius corresponding to the Fe II 5169 \AA\ line and therefore use this feature to derive the velocity of the photosphere, estimating a 15\% uncertainty. As acknowledged in \cite{Ergon2014}, the Fe II 5169 \AA\ line is formed above the photosphere and while it can be a good estimator of the photosphere velocity in SN~II-P \citep{Dessart2005}, it is still unclear how this translates to SESNe. We therefore favor the velocity associated with the blackbody emitting surface, which is derived from the photometric evolution presented in \cite{Ergon2014} (see Figure~14 within). The bolometric light curve peaks at about 21 days after the explosion epoch, so we adopt $v_{\rm ph}(t_{\rm rise}) \approx 4.64 \times 10^8 \ {\rm cm}\,{\rm s}^{-1}$.  

In \cite{Ergon2015}, the authors utilize hydrodynamical modeling out to 400 days to fit the bolometric light curve and find good agreement with the explosion parameters determined in previous work, though offering a new constraint of the helium core mass, $M_{\rm He} = 3.06_{-0.44}^{+0.68}\, M_{\odot}$. Given the iron core masses of our models in a similar ZAMS mass range shown in Figure~\ref{fig:masses}, we adopt a conservative value of $M_{\rm Fe} = 1.45\, M_{\odot}$. This gives us a lower limit ejecta mass of $M_{\rm ej} = 1.61_{-0.44}^{+0.68}\, M_{\odot}$.

\begin{figure}
\includegraphics[width=0.47\textwidth,trim=0.0cm 0.0cm 0.5cm 0.0cm, clip]{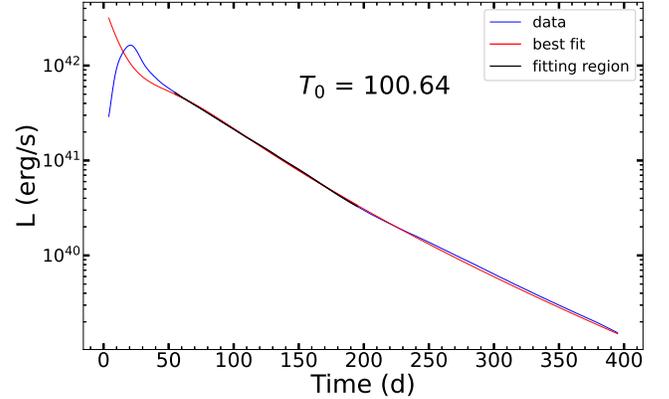}
\caption{Our fit of the bolometric light curve of SN 2011dh. As in Figure~\ref{fig:fit}, the data is shown in blue, the best fit modified heating equation is shown in red, and the black line indicates the fitting region from 57 \textemdash 200 days. We find the data is best fit by $T_0 = 100.64$ days. }
\label{fig:2011dh}
\end{figure}

Using our fitting procedure described in Section~\ref{sec:models}, we fit the bolometric light curve of SN 2011dn and find a best fit $T_0 = 100.64$ days, shown in Figure~\ref{fig:2011dh}. Using the calibrated relationship between $T_0$, $M_{\rm ej}$, and $v_{\rm ph}$ stated in Equation~\ref{eq:T0_3}, we estimate SN 2011dh to have an ejecta mass of $M_{\rm ej} = 1.72 \pm 0.19\, M_{\odot}$, which is within $\sim 7\%$ of the accepted value. Using our analytic technique, we are able to quickly achieve a robust estimate of the ejecta mass of SN~2011dh and avoid more detailed hydrodynamical modeling.

\section{Conclusion}
\label{sec:conclusion}

We have presented an improved method for estimating ejecta mass from late time bolometric light curves based on the framework provided by \cite{Colgate1980a} and \cite{wheeler2015}. Traditional methods of estimating ejecta masses from the rise time of the bolometric light curve require a number of assumptions about the physics that impacts the early light curve shape. The implications of these assumptions have been previously noted in the many works cited throughout and we have provided a more detailed exploration and discussion of them for completeness. In particular, the opacity during the rise can be strongly dependent on density, temperature, composition, and the photon frequencies. We have also shown that the nickel mixing and the treatment of recombination can strongly impact the rise of the SN and make it difficult to reliably derive an ejecta mass. We have shown it is possible to avoid those assumptions by utilizing the late-time phase of the light curve where the thermalization of the radioactive energy input is dominated by gamma-ray transport.

By fitting the slope of the decay tail of of a light curve to our modified heating equation (Equation~(\ref{eq:Ldecay})) over sufficiently late times, we can estimate $T_0$, the characteristic timescale for gamma-ray escape. We then calibrated the relationship between $T_0$, ejecta mass, and photospheric velocity to our suite of 77 numerical models. The relationships follow closely to the power laws we expect from analytic arguments with small improvements to the scalings. Applying these results to Equation~(\ref{eq:log}) allowed us to recover ejecta mass estimates for our models that are within $20\%$ of the true value and when exploring assumptions related to $^{56}{\rm Ni}$ mixing and concentration on the late time light curve, we find similarly robust ejecta mass estimates, showing that the late time light curve is not heavily influenced by these parameters. In comparing to the event SN~2011dh, we are able to quickly recover a robust ejecta mass estimate in good agreement with values derived from more detailed numerical modeling.

This work demonstrates that in the future it will be important to follow SESNe for more than 100 days to provide reliable constraints on ejecta masses. Once obtained, such data will be crucial for better understanding the difference between classes of SESNe, and whether they have a similar set of massive star progenitors. One issue that we have not addressed in our work is how to construct bolometric light curves for comparisons to our model. Many SESNe observations are missing infrared data that is an important contribution to the late light curve. Further work, building off of bolometric corrections derived in \cite{Bersten2009}, which calibrates corrections for $\sim 100$ days, is needed to generate a set of reliable late-time corrections to allow forthcoming optical data sets (e.g., the Vera Rubin Telescope) to be easily utilized with our results.  


\acknowledgments

We thank Tim Morton for sharing insight and suggestions regarding the statistics of our sample. We also thank Viktoriya Morozova, Maria Drout, Ashley Villar, Amir Sharon, and Jenna Stelmar for helpful conversations and feedback on previous drafts of this manuscript.

A.H. acknowledges support from the USC-Carnegie fellowship.

\bibliographystyle{yahapj}
\bibliography{bhrefs}

\end{document}